\documentclass[aps,prl,preprint,superscriptaddress,showpacs]{revtex4}

\usepackage{graphicx}

\begin{document}

\title{Time-Domain Measurement of Spontaneous Vibrational Decay of
  Magnetically Trapped NH}

\author{Wesley C. Campbell}
\email[]{wes@cua.harvard.edu}
\affiliation{Department of Physics, Harvard University, Cambridge,
  Massachusetts 02138, USA}
\affiliation{Harvard-MIT Center for Ultracold Atoms, Cambridge,
  Massachusetts 02138, USA}
\author{Gerrit C. Groenenboom}
\affiliation{Theoretical Chemistry, Institute for
  Molecules and Materials (IMM), Radboud University Nijmegen,
  Toernooiveld 1, 6525 ED Nijmegen, The Netherlands}
\author{Hsin-I Lu}
\affiliation{School of Engineering and Applied Sciences, Harvard
  University, Cambridge, MA 02138 USA}
\affiliation{Harvard-MIT Center for Ultracold Atoms, Cambridge,
  Massachusetts 02138, USA}
\author{Edem Tsikata}
\affiliation{Department of Physics, Harvard University, Cambridge,
  Massachusetts 02138, USA}
\affiliation{Harvard-MIT Center for Ultracold Atoms, Cambridge,
  Massachusetts 02138, USA}
\author{John M. Doyle}
\affiliation{Harvard-MIT Center for Ultracold Atoms, Cambridge,
  Massachusetts 02138, USA}
\affiliation{Department of Physics, Harvard University, Cambridge,
  Massachusetts 02138, USA}

\date{\today}

\begin{abstract}
The $v=1 \rightarrow 0$ radiative lifetime of NH ($X^3\Sigma^-,
v=1,N=0$) is determined to be $\tau_{\mathit{rad,}\mathrm{ exp.}} =
37.0 \pm 0.5_{\mathrm{stat}} {+2.0 \atop -0.8} {}_{\mathrm{sys}}
\mbox{ ms}$, corresponding to a transition dipole moment of
$\left|\mu_{10}\right| = 0.0540 {+0.0009 \atop -0.0018} \mbox{ D}$.
To achieve the long observation times necessary for direct time-domain
measurement, vibrationally excited NH ($X^3\Sigma^-, v=1,N=0$)
radicals are magnetically trapped using helium buffer-gas loading.
Simultaneous trapping and lifetime measurement of both the
NH($v=1,N=0$) and NH($v=0,N=0$) populations allows for accurate
extraction of $\tau_{\mathit{rad,}\mathrm{ exp.}}$.  Background helium
atoms are present during our measurement of
$\tau_{\mathit{rad,}\mathrm{ exp.}}$, and the rate constant for helium
atom induced collisional quenching of NH($v=1,N=0$) was determined to
be $k_{v=1} < 3.9\times10^{-15}\mbox{ cm}^3\mbox{s}^{-1}$.  This bound
on $k_{v=1}$ yields the quoted systematic uncertainty on
$\tau_{\mathit{rad,}\mathrm{ exp.}}$.  Using an \textit{ab initio}
dipole moment function and an RKR potential, we also determine a
theoretical value of $36.99 \mbox{ ms}$, in agreement with our
experimental value.  Our results provide an independent determination
of $\tau_{\mathit{rad},10}$, test molecular theory, and furthermore
demonstrate the efficacy of buffer-gas loading and trapping in
determining metastable radiative and collisional lifetimes.

\end{abstract}


\maketitle

Measurement of molecular abundances in astrophysical, atmospheric, and
chemical settings is critical to the discovery of new physics.  A
myriad of techniques are used to accomplish this.  One in particular,
infrared absorption spectroscopy, studies light passing through a
sample (\textit{e.g.} an atmosphere or stellar cloud) to determine
total absorption.  In combination with transition strengths measured
in the lab, molecular density can be determined.  However, measurement
of transition strengths for rovibrational transitions of radicals is
particularly difficult \cite{ChackerianJQSRT88}.  This is generally
due to imprecise knowledge of the radical density.  One approach that
removes the need for absolute density calibration is time domain or
linewidth measurement of the transition lifetime.  Despite the
intrinsic advantages of this approach, it has been seldom used for
weak transitions.  This is because measurement of lifetimes longer
than a few milliseconds is difficult; the long measurement time
required can be limited by collisional quenching or transit time.  A
by-product of these experimental limitations has been the increasing
importance of \emph{ab initio} calculations in determining transition
strengths \cite{LievinPS92}.

One ``striking'' example of the need for improved experimental methods
and verification of \textit{ab initio} techniques is the case of
imidogen (NH).  The NH vibrational radiative lifetime
$\tau_{\mathit{rad}}$ is a vital number for astrophysical studies
since the $\Delta v=1$ transitions of the $X^3\Sigma^-$ state fall in
the 3 $\mu$m atmospheric transmission window.  The absolute line
strength of the 0-1 vibration-rotation (V-R) transition ($A_{10}$)
plays a crucial role in determining the nitrogen abundance in cool
stars
\cite{LambertApJS86,LambertApJ86,LambertApJ84,LambertPrivateCommunication},
which, in turn, is tied to questions of stellar evolution and internal
mixing.  The estimated absolute nitrogen abundance of the sun is based
on atomic nitrogen (N{\scshape i}) lines and NH V-R absorption
\cite{AsplundSSR07}.  The lack of precise knowledge of $A_{10}$ is the
dominant source of uncertainty in observed abundances based on V-R
lines \cite{GrevesseAA90}.  Despite their importance, measurements of
the NH V-R transition strengths are particularly lacking
\cite{BeerApJ72,DoddJCP91,GrevesseAA90}.

Chackerian, Jr. \emph{et al.}  \cite{ChackerianJCP89} have made the
only published laboratory measurements of $\tau_{\mathit{rad}}$ for NH
by observing the intensity variation introduced by the
vibration-rotation interaction (the Herman-Wallis effect
\cite{HermanWallisJCP55}) and combining this with the only measurement
of the permanent electric dipole moment \cite{DalbyCJP74} of
NH($X^3\Sigma^-$).  This experimental determination and the published
theoretical values
\cite{WernerJMS80,DoddJCP91,GrevesseAA90,LievinPS92,StevensJCP74}
cover a range spanning nearly 50\%.

In this Letter we report a direct time-domain measurement and
\textit{ab initio} theoretical calculation of the spontaneous emission
lifetime of vibrationally excited NH($v=1,N=0$) radicals.  NH ($v=1$
and $v=0$) is loaded into a magnetic trap using $^3$He buffer-gas
loading.  The trap lifetime of NH($v=1$) is measured to be
$\tau_{\mathit{trap,}v=1} = 33.5 \pm 0.5 \mbox{ ms}$, dominated by the
radiative lifetime ($\tau_{\mathit{rad,}10}$).  Corrections to the
measured lifetime due to the finite trap evaporation (determined from
our measured NH($v=0$) lifetime) and the possibility of background
helium collisional quenching ($k_{v=1}$) result in
$\tau_{\mathit{rad,}\mathrm{ exp.}} = 37.0 \pm 0.5_{\mathrm{stat}}{
+2.0 \atop -0.8} {}_{\mathrm{sys}} \mbox{ ms}$, in agreement with our
calculated value of $36.99\mbox{ ms}$.  We also measure a tight bound
on the helium-induced NH($v=1$) collisional quenching rate coefficient
of $k_{v=1} < 3.9 \times 10^{-15}\mbox{ cm}^3{s}^{-1}$.

In 1998 our group demonstrated the first magnetic trapping of
molecules \cite{DoyleNATURE98} by buffer-gas loading CaH($v=0$).  It
was discovered that both $v=0$ and $v=1$ states were coolable and
CaH($v=1$) persisted in the buffer gas as long as CaH($v=0$)
\cite{JonathanThesis}, suggesting that the collision-induced quenching
of vibrational states in the buffer gas was slow enough to utilize
buffer-gas loaded molecules to measure vibrational decay lifetimes.
In 2005, the Meijer group in Berlin used a very different method,
Stark-deceleration \cite{MeijerPRL99} and optical pumping, to
electrostatically trap OH($v=1$) and perform the first vibrational
lifetime measurement of cold trapped molecules \cite{MeijerPRL05}.
Their measurement improved the precision of the experimentally
determined Einstein A coefficient $A_{10}$ to better than 4\% and was
in agreement with their \textit{ab initio} calculated value.  Our
group recently demonstrated trapping of $^3\Sigma$ state molecules
with the buffer-gas loading of ground-state NH into a magnetic trap
\cite{DoylePRL07}.  The large splitting between levels of opposite
parity in molecules in $\Sigma$ states makes them unsuitable for Stark
deceleration due to nonlinear Stark shifts \cite{MeijerPRL99} but well
suited for buffer-gas loading.  We find that unlike CaH, the lifetime
of the vibrationally excited NH molecules in the trap is significantly
shorter than the ground state due to spontaneous emission.

The experimental apparatus and method is essentially the same as that
described in Ref. \cite{DoylePRL07}.  Briefly, NH radicals are loaded
into a magnetic trap from a molecular beam.  About $\approx 10$\% of
the NH produced in the beam by dissociating ammonia in a DC glow
discharge is in the $v=1$ state.  We do not detect any NH($v=2$).  The
NH($v=1$) along with NH($v=0$) enter the magnetic trap, which is
inside a cold buffer-gas cell.  The cold cell has a 1 cm diameter
entrance aperture for loading the NH radicals from the beam so that
the buffer gas exits through the aperture but is replenished through a
fill line.  The flow rate of buffer gas into the cell is controlled to
set the buffer-gas density at a constant level.  Low-field seeking
molecules fall into the 3.9 Tesla magnetic trap and are detected using
either laser induced fluorescence (LIF) or laser absorption on the $A
\leftrightarrow X$ transition.  Laser absorption spectroscopy
indicates that we typically trap more than $2\times 10^8$ NH($v=0$)
molecules.  The excitation laser is a 672 nm CW dye laser frequency
doubled to 336 nm \footnote{LD688 dye in EPH, frequency doubled in BBO
in an external buildup cavity}.  Thermalization of the external and
rotational motion of the NH proceeds rapidly, as can be verified from
their trapped lifetime and the Zeeman-broadened spectrum
\cite{DoylePRL07}.

\begin{figure}
\includegraphics[width=3in]{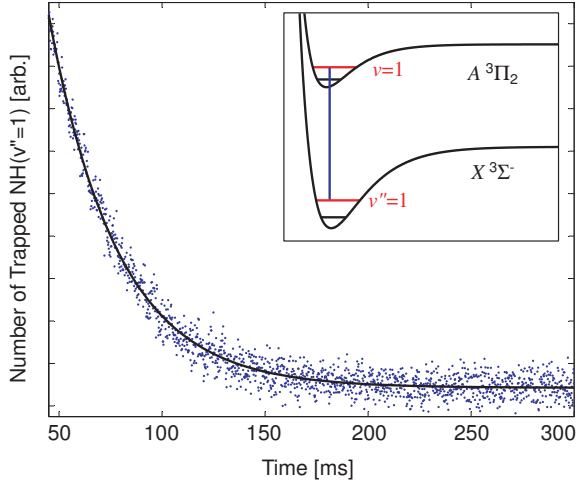}
\caption{Vibrationally-excited trapped molecule density vs. time.  The
  population of the NH $X^3\Sigma^{-}(v^{\prime\prime}=1, N=0)$
  molecules is monitored using laser-induced fluorescence on the 1-1
  band of the $A \leftrightarrow X$ transition (see inset), and the
  time profiles are fitted to a single-exponential decay to obtain
  observed lifetimes.  The loading process that occurs in the first 50
  ms is essentially the same as described in \cite{DoylePRL07}, and we
  begin fitting as least 5 ms after the molecular beam has been turned
  off to ensure rotational and translational themalization.\label{TimeProfile}}
\end{figure}

\begin{figure}
\includegraphics[width=3in]{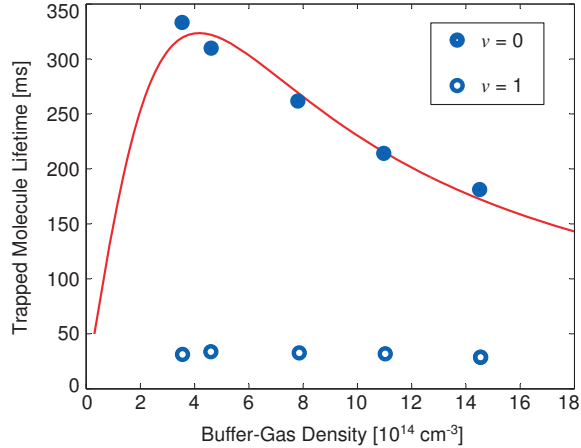}
\caption{Measured trap lifetime for NH($v=0$) (closed circles,
  $\tau_{\mathrm{hold}}$) and NH($v=1$) (open circles,
  $\tau_{\mathrm{trap,}v=1}$).  For the NH($v=0$), the lifetime is set
  by a combination of the trap depth, the diffusion enhancement of the
  lifetime, and collision-induced Zeeman relaxation.  The solid curve
  is a fit to the expected functional form \cite{DoylePRL07}.  The
  lifetime of NH($v=1$) is limited by the spontaneous emission
  lifetime.\label{TauvsN}}
\end{figure}

NH($v=0$) and NH($v=1$) are detected from the rotational ground state
($N=0$) on the diagonal vibrational lines of the $\mathrm{R}_1 \: A
\leftrightarrow X$ transition.  Figure \ref{TimeProfile} shows a
measurement of the number of trapped NH($v=1$) versus time,
\textit{i.e.} the fluorescence observed due to our continuous
excitation of the $A \leftarrow X$ transition.  Lifetimes are measured
by $\chi^2$-fitting the fluorescence to a single exponential and the
measured lifetime does not depend on laser power in this range.  The
$\chi^2$ fits are derived from error bars empirically determined from
repeated LIF counting measurements and calculated shot noise.  Figure
\ref{TauvsN} shows the measured trap lifetime of both NH($v=1$) and
NH($v=0$) vs. buffer-gas density at a temperature of 615 mK.  The
vibrational ground state lifetime shows the previously studied
dependence on buffer-gas density \cite{DoylePRL07}.  At low buffer-gas
densities, the helium enhances the trap lifetime by enforcing
diffusive motion of the trapped molecules.  As the buffer-gas density
is increased further, the lifetime is suppressed due to the higher
frequency of Zeeman relaxation collisions with helium.  In contrast,
the measured lifetime of the $v=1$ state ($\tau_{\mathrm{trap,}v=1}$)
shows very little buffer-gas density dependence, as is expected for
spontaneous radiative decay.

\begin{figure}
\includegraphics[width = 3in]{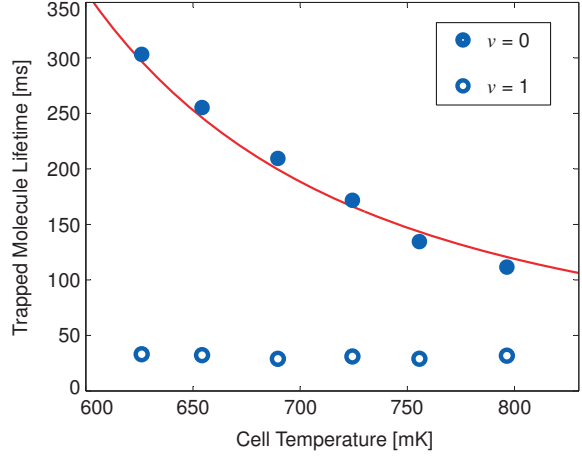}
\caption{Measured trap lifetimes for the ground (closed circles) and
  first vibrationally excited states (open circles) of NH
  vs. buffer-gas temperature.  The $v=0$ molecules show the previously
  studied strong dependence on temperature \cite{DoylePRL07}.  The
  solid curve is a fit to the expected functional form.  The $v=1$
  state shows very little dependence on buffer-gas
  temperature.\label{TauvsT}}
\end{figure}

Figure \ref{TauvsT} shows trap lifetimes for NH($v=0$) and NH($v=1$)
as the temperature of the buffer gas is varied.  The NH($v=0$) data
show the expected strong dependence of lifetime on trap depth, as
described in Ref. \cite{DoylePRL07}.  In contrast, the NH($v=1$) trap
lifetime changes by less than 5 ms over this range, which again
demonstrates that the lifetime is not dominated by either trap depth
or collisions.  This is again consistent with spontaneous radiative decay.

The spontaneous radiative lifetime of NH($v=1$) is measured
using the following procedure.  We repeatedly measure the lifetime of
trapped NH($v=1$) at $615 \mbox{ mK}$ with buffer-gas density
$n_{\mathrm{He}} = 3.6 \times 10^{14} \mbox{ cm}^{-3}$.  This
measurement is done at the low end of our workable buffer-gas
density range, resulting in relatively long lifetimes for trapped NH($v=0$),
thereby reducing the effect of finite trap lifetime on the measured
value of $\tau_{\mathit{rad,}\mathrm{ exp.}}$.  Data is obtained from 1041
trapping cycles and fluorescence intensity time profiles taken over
four hours, yielding a measured NH($v=1$) trap lifetime of
$\tau_{\mathrm{trap,}v=1} = 33.5 \mbox{ ms}$ with a standard error of
$\pm 0.5 \mbox{ ms}$.

In order to extract the $1 \rightarrow 0$ spontaneous radiative
lifetime $(\tau_{\mathit{rad,}\mathrm{ exp.}})$ from the measured trap lifetime
$(\tau_{\mathrm{trap,}v=1})$, the finite hold time of the trap
$(\tau_{\mathrm{hold}})$ as well as additional loss mechanisms unique to the
vibrationally excited molecules $(\tau_{\mathrm{loss}})$ must be taken
into account.  $\tau_{\mathit{rad,}\mathrm{ exp.}}$ can be written

\begin{equation}
\tau_{\mathit{rad,}\mathrm{ exp.}} = \frac{1}{\tau_{\mathrm{trap,}v=1}^{-1} -
\tau_{\mathrm{hold}}^{-1} - \tau_{\mathrm{loss}}^{-1}}.
\label{tau_relations}
\end{equation}
The finite trap hold time effects are measured directly by observing
the trap lifetime of NH($v=0$).  Given that the magnetic moment of
NH($v=1$) is identical to NH($v=0$), this gives a simple correction of
$\Delta \tau = +3.5 \pm 0.1 \mbox{ ms}$.

Among the possible mechanisms that would cause population changes in
NH($v=1$), background helium collision-induced loss is the only likely
mechanism we can identify for our system.  We will refer to this loss
mechanism as vibrational quenching, although it should be noted that
if the collision-induced Zeeman relaxation for the $v=1$ state is
different from the $v=0$ state, it would manifest itself with the same
functional dependence on $n_{\mathrm{He}}$ as direct $v=1 \rightarrow
0$ quenching.  In order to quantify this possible effect, we fit the
data in Fig. \ref{TauvsN} assuming as a free parameter helium
collision-induced loss of NH($v=1$) (after correcting
$\tau_{\mathrm{trap,}v=1}$ for the trap hold time).  This attempted
fit results in a collisional-quenching rate coefficient of $1.1 \pm
2.8 \times 10^{-15} \mbox{ cm}^3\mbox{s}^{-1}$, consistent with zero.
This implies an upper limit on the collisional quenching rate
coefficient for NH($v=1$) of $k_{v=1} < 3.9 \times 10^{-15} \mbox{
cm}^3\mbox{s}^{-1}$ and a consequent systematic correction of the
lifetime from this possible effect of between $-0.8$ and $+2.0 \mbox{
ms}$ where the negative value would correspond to slower Zeeman
relaxation for $v=1$ than $v=0$.  This is the dominant systematic
uncertainty in our measurement, which yields a final value of
$\tau_{\mathit{rad,}\mathrm{ exp.}} = 37.0 \pm
0.5_{\mathrm{stat}}{+2.0 \atop -0.8} {}_{\mathrm{sys}} \mbox{ ms}$
where the second uncertainty is the systematic uncertainty (completely
due to our measured limit on the collision-induced quenching rate
coefficient).  Fractional corrections to the lifetime from blackbody radiation
\cite{MeijerPRL07}, hyperfine splitting \cite{RadfordMP76} and Zeeman
shifts \cite{RadfordMP76} are calculated to be negligible ($<10^{-4}$).

We can use this measurement to extract the dipole matrix element for
this transition, $\mu_{10} \equiv
\left<\phi_{v=1,N=0}|\mu|\phi_{v=0,N=1}\right>$ where $\phi_{v,N}$ is
the vibrational wavefunction.  The spontaneous emission lifetime can
be written in terms of the vibrational transition dipole moment as
\begin{equation}
\tau_{\mathit{rad},10} = \frac{3c^2e^2}{4\alpha \omega^3} \left| \langle \phi_{v=1,N=0}(r)|\mu(r) |\phi_{v=0,N=1}(r)
\rangle \right|^{-2}, \label{tau_exact}
\end{equation}
where $\alpha$ is the fine-structure constant and $\omega$ is the
transition frequency.  The final rotational state for the decay to
$v=0$ is $N=1$ due to the parity selection rule.  We calculate the
contribution to the lifetime from the $N=3$ final state to be
negligible ($<10^{-5}$).  Taking $\omega = 3092.88\mbox{ cm}^{-1}$
from spectroscopic constants in the literature
\cite{RamJCP99,RamCorrectionJCP99} gives us a measured value of
$\left|\mu_{10}\right| = 0.0540 {+0.0009 \atop -0.0018} \mbox{ D}$.

Comparison of our experimental value of $\tau_{\mathit{rad,
    }\mathrm{exp.}}$ to theory is crucial to claim resolution of the
    conflicting published measured and theoretical values.  To obtain
    a state-of-the-art calculated value we perform an \textit{ab
    initio} calculation of the spontaneous emission lifetime of
    NH($v=1$).  The vibrational wave function $\phi_{v=1,N=0}(r)$ was
    computed by solving the one-dimensional Schr\"odinger equation
    with the sinc-function discrete variable representation method
    \cite{GroenenboomJCP93}.  The radial potential was constructed
    with the RKR method \cite{LeRoy04}, using the Dunham parameters
    from Ram {\em et al.}  \cite{RamJCP99,RamCorrectionJCP99}.  To
    compute $\phi_{v=0,N=1}(r)$ the centrifugal term was added to the
    potential.  The {\sc molpro} program package \cite{molpro} was
    used to compute the $r$-dependent electric dipole moment function
    $\mu(r)$. We used the internally contracted multireference single
    and double excitation configuration interaction (MRCI) method
    \cite{WernerJCP88} and the aug-cc-pV6Z one electron basis.  The
    orbitals were obtained in a complete active space self consistent
    field (CASSCF) calculation. The active space consisted of the
    $1\sigma-6\sigma$, $1\pi-3\pi$, and $1\delta-2\delta$
    orbitals. All CASSCF configurations were used as reference
    configurations in the MRCI calculation.  The result is a
    calculated spontaneous emission lifetime of 36.99 ms, in excellent
    agreement with our measured value.

To get an indication of the accuracy of the \textit{ab initio}
calculation we also computed the spectroscopic constants and the
lifetime using the MRCI potential instead of the RKR potential. This
results in a transition frequency that is 0.44\% too high, a ground
state rotational constant that is 0.48\% too high and a computed
lifetime of 37.29 ms.

Another indication of the accuracy of the \textit{ab initio}
calculation was obtained by computing the dipole function with the
partially spin restricted coupled cluster method with single and
double excitations and a perturbative treatment of the triples
(RCCSD(T)) \cite{WernerJCP93,WernerErratumJCP00,BartlettJCP93}. This
is a single reference method and the orbitals were taken from a
restricted Hartree Fock calculation. Core-excitations were included
and the aug-cc-pV6Z basis was used. The dipole was obtained in a
finite field calculation with electric fields of $\pm 10^{-4}$
$E_h/a_0$.  With the vibrational wave functions computed for the RKR
potential this results in a lifetime of 37.39 ms. The transition
frequency computed with the RCCSD(T) potential is 0.73\% too high.

Using the MRCI dipole moment function we also computed the NH($v=0$)
\textit{permanent} dipole moment.  The result, 1.5246 D, is 10\%
larger than the 1974 experimental value of 1.38$\pm$0.07 D
\cite{DalbyCJP74} but it agrees well with the 1996 \textit{ab initio}
value of 1.536 D Paldus and Li \cite{LiCJC96}, who already express doubt
about the experimental value.

While both our measured and calculated values for the transition
dipole moment agree, they differ significantly from previously
published values.  Chackerian, Jr. \textit{et al.}
\cite{ChackerianJCP89} have observed the Herman-Wallis effect and
combined their spectroscopic data with the 1974 measured value of the
static dipole moment of ground-state NH \cite{DalbyCJP74}.  This
results in a measured value for the transition dipole moment of
$\left|\mu_{10}\right| = 0.0648 \pm 0.008 \mbox{ D}$, which they
present with a calculated \textit{ab initio} value
\cite{ChackerianJCP89} of $\left|\mu_{10}\right| =0.0594 \mbox{ D}$.
These transition dipole moments are inconsistent with our result of
$\left|\mu_{10}\right| = 0.0540 {+0.0009 \atop -0.0018} \mbox{ D}$.
Grevesse \textit{et al.}  provide a plot of the calculated dipole
moment matrix element (including rotation) vs. $N^{\prime\prime}$ that
gives a value of $\left|\mu_{10}\right| > 0.06\mbox{ D}$ for
$N^{\prime\prime}=1$ \cite{GrevesseAA90}.  Cantarella \textit{et al.}
use a series of different published potentials and dipole moment
functions to calculate dipole matrix elements falling in the range of
$0.0528 \le \left|\mu_{10}\right| \le 0.0618\mbox{ D}$
\cite{LievinPS92}.  Das \textit{et al.} calculate and oscillator
strength corresponding to $\left|\mu_{10}\right| = 0.0526\mbox{ D}$
\cite{StevensJCP74}.

Our measured lifetime $\tau_{\mathit{rad,}\mathrm{ exp.}} = 37.0 \pm
0.5_{\mathrm{stat}} {+2.0 \atop -0.8} {}_{\mathrm{sys}} \mbox{ ms}$
can also be compared to published values of the Einstein A coefficient
using $A_{10}^{-1} = \tau_{\mathit{rad,}10}$.  Rosmus and Werner
\cite{WernerJMS80} calculate $A_{10}^{-1} = 28.7 \pm 4 \mbox{ ms}$,
and Dodd \textit{et al.}  \cite{DoddJCP91} calculate $A_{10}^{-1} =
19.3 \mbox{ ms}$.  Both of these calculations neglect rotation, and we
find convergence for our calculated value only after inclusion of the
large basis and active space described above.  Despite these
discrepancies, our measurement is in agreement with our \textit{ab
  initio} calculated value and we believe this new value to be the
most accurate yet reported.

In conclusion, we have measured the spontaneous emission lifetime of
the $X^3\Sigma^{-}(v=1,N=0)$ state of the NH radical in a magnetic
trap and performed a new \textit{ab initio} calculation and find these
in good agreement.  This new value can be used to calibrate molecular
densities in spectroscopic measurements on the NH $1-0$ transition.
Since previously published values for the transition dipole matrix
element are almost all larger than our measured value, it is likely
that NH densities that have been inferred from V-R absorption on this
line are lower than the actual density of NH radicals.  We estimate
that current solar nitrogen abundance estimates derived from this
transition \cite{GrevesseAA90,AsplundConference,AsplundSSR07} will
require a correction of about 30\%.

Using buffer-gas loading at low helium densities, we have shown the
effect of the buffer gas to be small enough to measure the lifetime of
$\tau_{\mathit{rad,}10}[\mbox{NH}]$ in a magnetic trap with the
highest precision yet reported.  Since the Zeeman relaxation cross
section for $v=1$ and $v=0$ are unlikely to be very different, future
measurement of the vibrational quenching cross section of NH($v=1$)
with helium may reduce the systematic error bars of this measurement
to less than our current statistical uncertainty.  Our work
demonstrates a new application for buffer-gas loading that can benefit
the physics, astrophysics, and chemistry communities.

\begin{acknowledgments}
The authors would like to thank Nathan Brahms, Amar Vutha, and David
Lambert for helpful discussions, as well as Kevin Strecker for
suggesting LD688 laser dye.  This work was supported by the
U.S. Department of Energy under Contract No. DE-FG02-02ER15316 and the
U.S. Army Research Office.
\end{acknowledgments}

\bibliography{Wesbib.bib}

\begin{thebibliography}{33}
\bibitem[{Ram()}]{RamCorrectionJCP99}
\bibinfo{note}{To reproduce the term values in Table I of \cite{RamJCP99} we
  had to assume this sign convention in Table II: $B_e=Y_{0,1}$,
  $\alpha_e=-Y_{1,1}$, $\gamma_e = Y_{2,1}$, $\delta_e=-Y{3,1}$, and
  $\epsilon_e=-Y_{4,1}$.}
\end{thebibliography}

\end{document}